\documentclass[a4paper,twocolumn]{esapub}
\usepackage{times}
\usepackage{epsf,graphicx}
\usepackage{natbib}

\def\mbf#1{\mbox{\boldmath ${#1}$}}
\def\Alfven{Alfv\'{e}n~}

\def\lesssim{\; \buildrel < \over \sim \;}
\def\gtrsim{\; \buildrel > \over \sim \;}
\def\apj{ApJ}
\def\araa{ARA\&A}
\def\aap{A\&A}
\def\jgr{JGR}
\def\apjl{ApJL}
\def\ssr{Spac. Sci. Rev.}
\def\nat{Nature}
\def\grl{Geophys. Res. Lett.}
\def\aaps{A\&AS}
\def\solphys{Sol.Phys.}
\def\mnras{MNRAS}

\title{Successful Coronal Heating and Solar Wind Acceleration 
by MHD Waves by Numerical Simulations from Photosphere to 0.3AU}

\author{Takeru K. Suzuki\footnote{JSPS Research Fellow} \& Shu-ichiro Inutsuka}
\affil{Department of Physics, Kyoto University, Kitashirakawa, 
Kyoto, 606-8502, Japan; stakeru@scphys.kyoto-u.ac.jp}

\begin{document}

\maketitle
\begin{abstract}
We show that the coronal heating and the acceleration of the fast solar wind 
in the coronal holes are natural consequence of the footpoint fluctuations 
of the magnetic fields at the photosphere   
by one-dimensional, time-dependent, and nonlinear magnetohydrodynamical 
simulation with radiative cooling and thermal conduction.  
We impose low-frequency ($<0.05$Hz) transverse photospheric motions, 
corresponding to the granulations, with velocity 
$\langle dv_{\perp}\rangle = 0.7$km/s. 
In spite of the attenuation in the chromosphere by the reflection, 
the sufficient energy of the generated outgoing \Alfven waves transmit into 
the corona to heat and accelerate of the plasma by nonlinear 
dissipation. 
Our result clearly shows that the initial cool ($10^4$K) and static 
atmosphere is naturally heated up to $10^6$K and accelerated to 
$\simeq 800$km/s, and explain recent SoHO observations and Interplanetary 
Scintillation measurements.
\end{abstract}

\section{Introduction}
It is still poorly understood how the coronal heating and the acceleration 
of the high-speed solar winds are accomplished in the open coronal holes which 
mainly exist in the polar regions except at the solar maximum phase. 
The origin of the energy to heat and accelerate the plasma is believed to be 
in the surface convection. This energy is lifted up 
through the magnetic fields and its dissipation leads to the plasma heating. 
In general, the problem of the coronal heating and the solar wind 
acceleration is to solve how the solar atmosphere reacts to the footpoint 
motions of the magnetic fields. 
Then, an ideal way to answer the problem is to solve the transfers of mass, 
momentum, and energy in a self-consistent manner from the photosphere to 
the interplanetary space, although such an attempt has not been successful 
yet.

In the coronal holes the \Alfven wave is regarded to play an important role 
(e.g. Belcher 1971), 
since it can travel a long distance to contribute not only to the coronal 
heating but to the solar wind acceleration. 
The \Alfven waves are excited by steady transverse motions of the field 
lines at the photosphere (e.g. Cranmer \& van 
Ballegooijen 2005), while they can also be produced by continual reconnections 
above the photosphere \citep{am97}. 
The latter process might be responsible for generation of the 
high-frequency (up to $10^4$Hz) ioncyclotron wave highlighted 
for the heating of the heavy ions \citep{kol98}.  
However, it is difficult to sufficiently heat the protons having the higher 
resonant frequency by the high-frequency waves, because the wave energy is 
already absorbed by the heavy ions with higher mass-to-charge ratio 
({\it i.e. lower} resonant frequency) \citep{cra00}.  
On the other hand, the low-frequency  
($\lesssim 0.1$Hz) \Alfven wave does not have such a drawback; it can 
propagate a long distance and heat the bulk of the plasma.  
Since we focus on the heating of the major part of the plasma, we study the 
low-frequency \Alfven waves by the steady footpoint 
fluctuations.

\begin{figure}
\includegraphics[width=1.\linewidth]{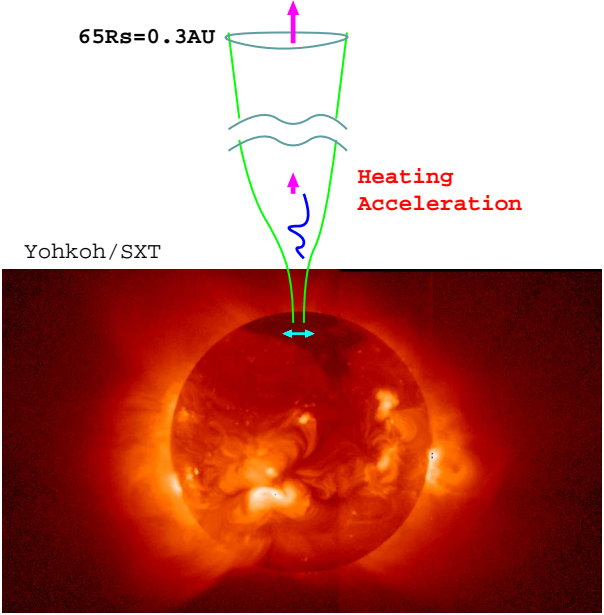}
\caption
{Schematic picture of our simulation (especially for NON-solar 
physicists). We consider the waves in the open flux tube rooted at the 
photosphere in the coronal hole to 0.3AU. 
Note that a coronal hole (dark region) is clearly seen in the 
north polar region in Yohkoh/SXT image.   
}
\label{fig:cartoon}
\end{figure}

The low-frequency \Alfven waves have been intensively studied in the context 
of the heating and acceleration of the solar wind plasma. 
For example, \citet{oug01} and \citet{dmi02} examined turbulent cascade 
of the \Alfven waves in tangential directions.  
Recently, \citet{ofm04} performed three-fluid and two-dimensional simulations 
for the wave-driven fast solar wind.  
These works are quite important in understanding the physical mechanism of 
the solar wind acceleration. 
One of the shortcomings of the previous research is that the most calculations 
for the wave-driven solar wind adopt 
the fixed 'coronal base', instead of the photosphere, as the inner boundary.  
In reality, however, the corona dynamically interacts with the chromosphere 
located below by chromospheric evaporation \citep{ham82} and wave 
transmission \citep{ks99} so that the coronal base varies time-dependently. 
Although some calculations \citep{lsv01,lsv02} include the chromosphere and 
the transition region to properly take into account these effects, they 
assume unspecified mechanical energy deposition which requires an ad hoc 
heating function with the phenomenological dissipation length.

In contrast, 
we self-consistently treat the transfer of the mass/momentum/energy by 
dynamically solving the wave propagation from the photosphere to the 
interplanetary region. 
We handle the heating and acceleration by the waves 
in an automatic way without the phenomenological heating function. 
Our aim is to answer the problem of the heating and acceleration in the 
coronal holes by the {\em forward} approach \citep{gn05} with the 
{\em minimal} unknown parameters.

\section{Simulation Method}

We consider one-dimensional open flux tube which is super-radially 
open (fig.\ref{fig:cartoon}), 
measured by heliocentric distance, $r$. 
The simulation region is from the photosphere ($r=1R_{\rm S}$) with  
density, $\rho = 10^{-7}$g cm$^{-3}$, to $65R_{\rm S}$ (0.3AU), 
where $R_{\rm S}$ is solar radius.  
Radial field strength, $B_r$, is given by conservation of magnetic flux as 
\begin{equation}
B_r r^2 f = {\rm const.} ,
\end{equation}
where $f$ is a super-radial expansion factor \citep{ko76}. 
In this paper $B_r$ is set to be $161$G at the photosphere, 
$5$G at low coronal height, $r=1.02 R_{\rm S}$, and 
$B_r = 2.1{\rm G}(R_{\rm S}/r)^2$ in $r>1.5 R_{\rm S}$ by controlling $f$. 
We give transverse fluctuations of the field line by the granulations at the 
photosphere, which excite \Alfven waves. 
We consider the fluctuations with power spectrum, $P(\nu)\propto \nu^{-1}$, 
in frequency between $6\times 10^{-4} \le \nu \le 0.05$Hz 
(period of 20seconds --- 30minutes), and root mean squared (rms) average 
amplitude $\langle dv_{\perp}\rangle\simeq 0.7$km/s corresponding to 
observed velocity amplitude $\sim 1$km/s \citep{hgr78}.
At the outer boundary, non-reflecting condition is imposed for all the MHD 
waves \citep{szi05a,szi05b}, which enables us to carry out 
simulations for a long time until quasi-steady state behaviors are achieved  
without unphysical wave reflection.

We dynamically treat propagation and dissipation of the waves and heating 
and acceleration of the plasma by solving 
ideal MHD equations with the relevant physical processes : 
\begin{equation}
\label{eq:mass}
\rho \frac{d}{dt}(\frac{1}{\rho}) - \frac{1}{r^2 f}\frac{\partial}{\partial r}
(r^2 f v_r ) = 0 , 
\end{equation}
$$
\hspace{-1cm}\rho \frac{d v_r}{dt} = -\frac{\partial p}{\partial r}  
- \frac{1}{8\pi r^2 f}\frac{\partial}{\partial r}  (r^2 f B_{\perp}^2)
$$
\begin{equation}
\label{eq:mom}
+ \frac{\rho v_{\perp}^2}{2r^2 f}\frac{\partial }{\partial r} (r^2 f)
-\rho \frac{G M_{\rm S}}{r^2}  , 
\end{equation}
\begin{equation}
\label{eq:moc1}
\rho \frac{d}{dt}(r\sqrt{f} v_{\perp}) = \frac{B_r}{4 \pi} \frac{\partial} 
{\partial r} (r \sqrt{f} B_{\perp}),
\end{equation}
$$
\hspace{-0.5cm}\rho \frac{d}{dt}(e + \frac{v^2}{2} + \frac{B^2}{8\pi\rho} 
-\frac{G M_{\odot}}{r}) 
+ \frac{1}{r^2 f} 
\frac{\partial}{\partial r}[r^2 f \{ (p + \frac{B^2}{8\pi}) v_r  
- \frac{B_r}{4\pi} (\mbf{B \cdot v})\}] 
$$
\begin{equation}
\label{eq:eng}
+ \frac{1}{r^2 f}\frac{\partial}{\partial r}(r^2 f F_{\rm c}) 
+ q_{\rm R} = 0,
\end{equation}
\begin{equation}
\label{eq:ct}
\frac{\partial B_{\perp}}{\partial t} = \frac{1}{r \sqrt{f}}
\frac{\partial}{\partial r} [r \sqrt{f} (v_{\perp} B_r - v_r B_{\perp})], 
\end{equation}
where $\rho$, $\mbf{v}$, $p$, $e$, $\mbf{B}$ are density, velocity, pressure, 
specific energy, and magnetic field strength, respectively, and subscript 
$r$ and $\perp$ denote radial and tangential components. 
$G$ and $M_{\rm S}$ are the 
gravitational constant and the solar mass. $F_{\rm c}$ is thermal 
conductive flux and $q_{\rm R}$ is radiative cooling 
\citep{LM90,aa89,mor04}. 
We adopt 2nd-order MHD-Godunov-MOCCT scheme for updating the physical 
quantities (Sano \& Inutsuka 2005),  
of which an advantage is that no artificial viscosity 
is required even for strong MHD shocks. 
We fix 14000 grid points with variable sizes in a way to resolve \Alfven waves 
with $\nu=0.05$Hz by at least $10$ grids per wavelength to reduce 
numerical damping. 

The advantages of our simulation are (i) automatically treating wave 
propagation/dissipation and plasma heating/acceleration without an ad hoc 
heating function 
(ii) with minimal assumption, {\it i.e.}, the transverse footpoint 
motions in the given flux tube (iii) in the broadest regions with respect 
to the density contrast amounting to 15 orders of magnitude from the 
photosphere to 0.3AU.    
On the other hand, the disadvantages are due to the 1-dimensional MHD 
approximation; we assume that the wave propagation is restricted to the 
direction parallel to the field line and the plasma behaves as the one-fluid 
with the Boltzmann particle distribution.  We will discuss this issues later.
We would like to emphasize however,  that this is the most self-consistent 
simulation for the solar wind acceleration, in spite of these shortcomings.

\begin{figure}
\includegraphics[width=1.\linewidth]{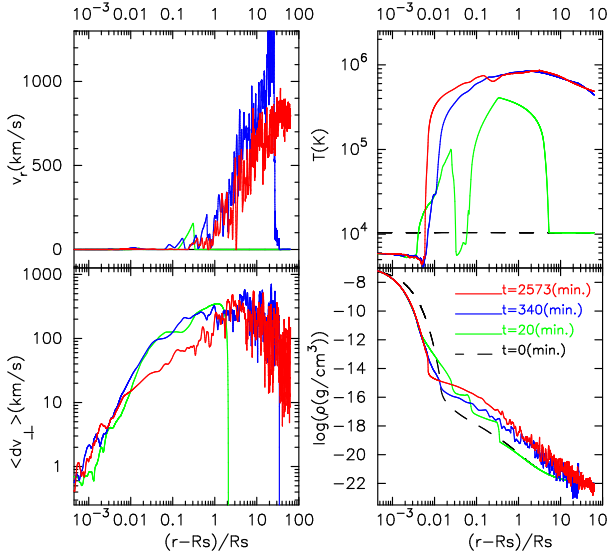}
\caption
{Time evolution of the atmosphere.  
Outflow speed, $v_r$(km/s) (upper-left), temperature, $T$(K) (upper-right),   
density, rms amplitude of transverse velocity, $\langle dv_{\perp}\rangle$
(km/s) (lower-left), and $\rho$(g/cm$^{3}$) (lower-right) are plotted. 
Black dashed, green, blue, and red solid lines are results at 
$t=0$, 20, 340, \& 2573 mins., respectively, whereas results 
at $t=0$ do not appear in $\langle dv_{\perp} \rangle$ 
because it equals to 0. 
The results are averaged by integrating for 3 minutes 
to take into account effects of the exposure time for comparison with 
observations.
}
\label{fig:tmevl}
\end{figure}

\section{Results}
We initially set static and cool atmosphere with temperature, $T=10^4$K; 
we do not impose the corona and the solar wind at the beginning.  
Figure.\ref{fig:tmevl}\footnote{Movie file is available :\\
http://www-tap.scphys.kyoto-u.ac.jp/\~{}stakeru/research/suzuki\_{}200506.mpg
} 
shows how the coronal heating and the 
solar wind acceleration are realized by the generated low-frequency \Alfven 
waves. 
We plot $v_r$(km/s), $T$(K), $\rho$(g/cm$^{3}$), 
and $\langle dv_{\perp}\rangle$(km/s)  
averaged from $v_{\perp}$ as a function of $(r-R_{\rm S})/R_{\rm S}$ 
at different time, $t=0, 20, 340$ \& 2573 minutes. 
As time goes on, the atmosphere is heated and accelerated 
effectively by dissipation of the \Alfven waves. 
Temperature rises rapidly in the inner region even at $t=20$minutes,    
and the outer region is eventually heated up by both outward 
thermal conduction and wave dissipation. 
Once the plasma is heated up to the coronal 
temperature, mass is supplied to the corona mainly by chromospheric 
evaporation due to the downward thermal conduction. 
This is seen in temperature structure as an inward shift of the transition 
region, which is finally located 
around $r =6 \times 10^{-3} R_{\rm S}$ ($\simeq$4000km), whereas it moves 
up and down by the time-dependent chromospheric evaporation (heating) and 
the wave transmission.  
As a result, the coronal density increases by two orders of magnitude. 
While the wind velocity exceeds 
1000km/s at $t=340$minutes on account of the 
initial low density, it gradually settles down to $< 1000$km/s  
as the density increases.   
$\langle dv_{\perp}\rangle$ also settles down to the reasonable value at the 
final stage. 
Temperature structure is smoother owing to the thermal 
conduction than the other quantities showing fluctuated behaviors 
due to the waves.

We have found that the plasma is steadily heated up to $10^6$K in the 
corona and flows out as transonic wind with $v_r \simeq 800$km/s at 
the outer boundary ($=$0.3AU) when the quasi 
steady-state behaviors are achieved after $t\gtrsim 1800$minutes.  
This is the first numerical simulation which directly shows that 
the heated plasma actually flows out as the transonic wind, initiated 
from the static and cool atmosphere, by the \Alfven waves. 
The outflow speed becomes up to $10$km/s even below $10^4$km 
above the photosphere ({\it c.f.} Tu et al.2005), although it is difficult 
to distinguish from the longitudinal wave motions. 
The sonic point where $v_r$ exceeds the local sound speed 
is located at $r\simeq 2.5R_{\rm S}$ and the \Alfven point 
is at $r\simeq 24R_{\rm S}$. 
Obtained proton flux at 0.3AU is  
$N_p v \simeq (2 \pm 0.5) \times 10^9$(cm$^{-2}$s$^{-1}$), corresponding to 
$N_p v \simeq (1.8 \pm 0.5) \times 10^8$(cm$^{-2}$s$^{-1}$) at 1AU 
for $N_p v \propto r^{-2}$, which is consistent with the 
observed high-speed stream around the earth \citep{apr01}. 

\begin{figure*}
\includegraphics[width=1.0\linewidth]{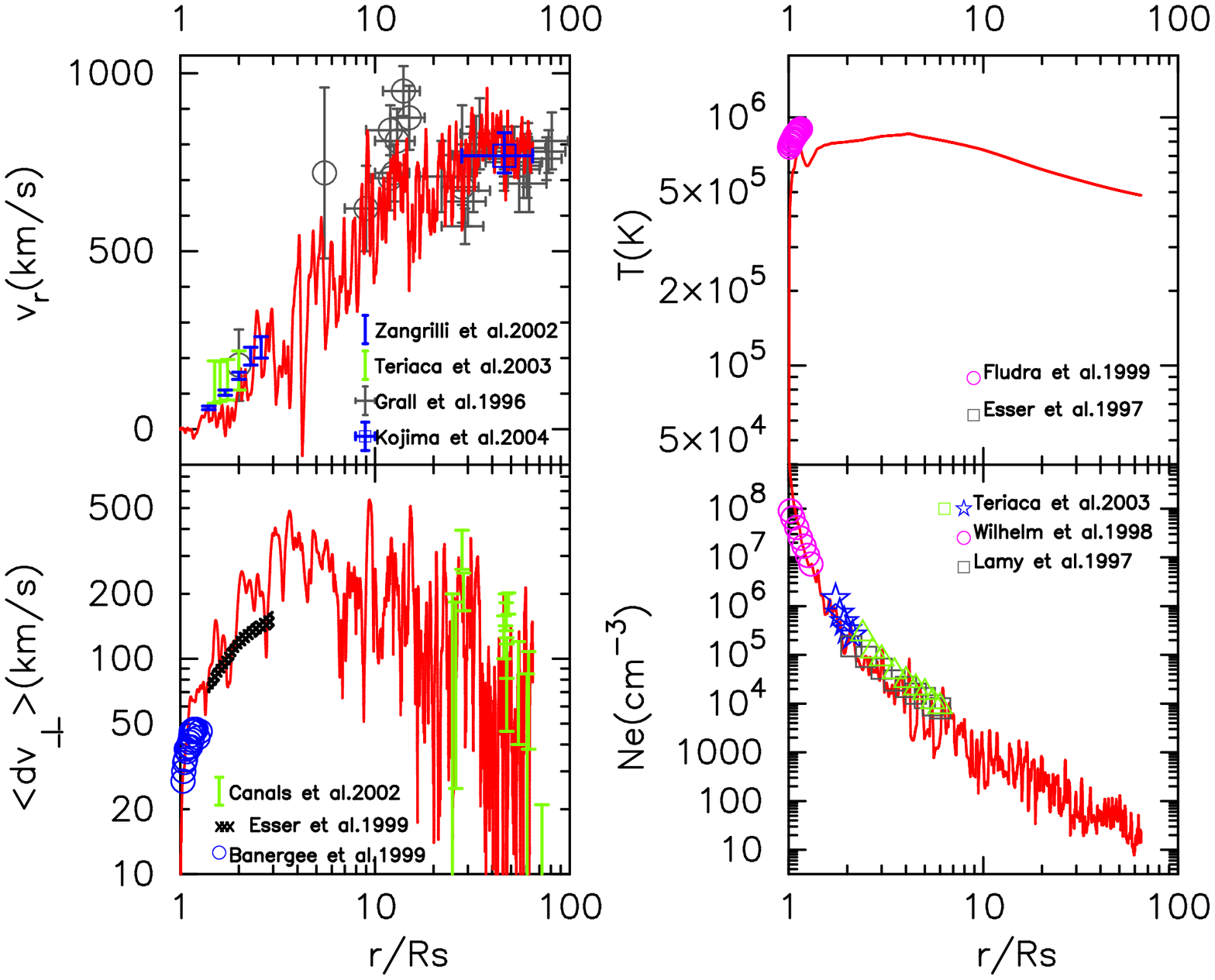}
\caption
{Comparison of the result at $t=2573$mins (red solid lines) 
with observations summarized below.  
Quantities in four panels are the same as in fig.\ref{fig:tmevl} except 
lower-right panel showing electron density, $N_e$(cm$^{-3}$), instead of 
$\rho$(g cm$^{-3}$). Scales 
in both horizontal and vertical axises are changed from fig.\ref{fig:tmevl}.  
{\bf a}: 
Vertical error bars with green x's \citep{tpr03} and blue triangles 
\citep{zan02} are proton outflow speed in polar regions by SoHO. 
Blue square with errors is velocity by IPS measurements 
averaged in 0.13 - 0.3AU of high-latitude regions\citep{koj04}. 
Crossed bars with and without circles are measurements by 
VLBI and IPS(EISCAT) \citep{gra96}.  
Vertical error bars with circles are data based on observation 
by SPARTAN 201-01 \citep{hab95}. 
{\bf b}: 
Pink circles are electron temperature by CDS/SoHO \citep{fdb99}. 
{\bf c}: Pink circles and blue stars are observations by SUMER/SoHO 
\citep{wil98} and by CDS/SoHO \citep{tpr03}, respectively. 
Green triangles \citep{tpr03} 
and squares \citep{lql97} 
are observations by LASCO/SoHO. 
{\bf d}: Blue circles are non-thermal broadening inferred from 
SUMER/SoHO measurements \citep{ban98}. 
Cross hatched region 
is empirically derived non-thermal broadening based on 
UVCS/SoHO observation \citep{ess99}. 
Green error bars with circles are transverse velocity 
fluctuations derived from IPS measurements by EISCAT \citep{can02}.         
}
\label{fig:obscmp}
\end{figure*}

In fig.\ref{fig:obscmp} we compare the result at $t=2573$minutes with 
recent observation in the high-speed solar winds from the polar regions 
by Solar \& Heliospheric Observatory (SoHO) 
\citep{zan02,tpr03,fdb99,lql97,wil98,ban98,ess99} 
and Interplanetary Scintillation (IPS) measurements \citep{gra96,koj04,can02}. 
The figure shows that our {\em forward} simulation naturally form the corona 
and the high-speed solar wind which are observed except small differences 
in detailed structures. 
Observed outflow speed in the inner corona($\le 3 R_{\rm S}$) 
and outer region($\gtrsim 20 R_{\rm S}$) 
are mostly explained by our simulation within the observed errors. 
Some of the observed data around $r \simeq 10R_{\rm S}$\citep{gra96} 
exceed our result, whereas it is reported that these data might reflect 
wave phenomena rather than the outflow speed \citep{hc05}. 
The simulated temperature shows a decent agreement with the electron 
temperature in the inner region by SoHO (Fludra et al.1997). 
Density 
and transverse amplitude 
show reasonable agreements with the observations. 

Our result manifestly shows that the heating and acceleration of 
the high-speed winds from the open field regions can be almost completely 
explained by the MHD dissipation mechanisms of the low-frequency \Alfven 
waves. 
The result is quite convincing since we automatically solve 
the transfers of mass/momentum/energy and the propagation 
of the \Alfven waves excited at the photosphere without any ad hoc 
assumptions.

\begin{figure}
\includegraphics[width=1.\linewidth]{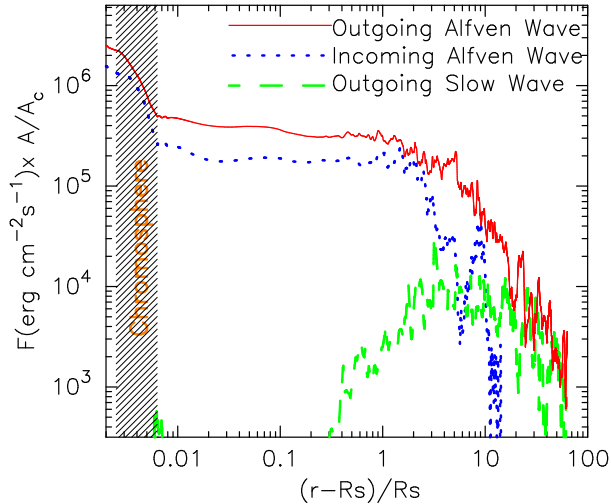}
\caption
{Energy flux of outgoing \Alfven (red solid), incoming \Alfven 
(blue dotted), and outgoing MHD slow (green dashed) waves at 
$t=2573$mins.  
Hatched region indicates the chromosphere. 
The energy flux is normalized by the cross section ($A=r^2 f$) of the flux 
tube at $r=1.02R_{\rm S}$ and $A_{\rm c}$ denotes 
$A$ at that location. 
}
\label{fig:engflx}
\end{figure}

Figure.\ref{fig:engflx} presents energy flux of outgoing \Alfven wave, 
incoming \Alfven wave, and outgoing MHD slow (sound) wave at $t=2573$minutes.  
To focus on the amount of dissipation the energy flux is normalized by 
the cross section of the flux tube at $r=1.02R_{\rm S}$ where it expands 
super-radially by 30 times from the photosphere. 
Note that the 
real energy flux in the lower (upper) regions is larger (smaller).  
The outgoing and incoming \Alfven waves are decomposed 
by correlation between $v_{\perp}$ and $B_{\perp}$ (Els\"asser variables). 
Extraction of the slow wave is also from fluctuating components of $v_r$ and 
$\rho$. 
The derived energy flux is averaged by integrating for 30 minutes 
to smooth out variation due to phase.

The figure shows that the outgoing \Alfven wave dissipate quite effectively; 
less than $10^{-3}$ of the initial energy is remained as that associated with 
the waves at the outer boundary. 
First, a sizable amount is reflected back downward 
below the coronal base ($r-R_{\rm S} < 0.01 
R_{\rm S} $), which is clearly 
illustrated in the energy flux of the incoming \Alfven wave following 
that of the outgoing component with slightly smaller level.  
This is because the wave shape is considerably 
deformed owing to the steep density gradient; 
a typical variation scale ($< 10^5$km) 
of the \Alfven speed becomes comparable or even shorter than 
the wavelength ($=10^4 - 10^6$km). 
Although the energy flux, $\simeq 5\times 10^{5}$erg cm$^{-2}$s$^{-1}$, 
of the outgoing \Alfven waves which penetrates into the corona is 
only $\simeq 15$\% of the input value,  
it meets the requirement for the energy budget in the coronal holes 
\citep{wn77}.

Second, MHD slow (sound) waves \citep{sak02} 
are generated in the corona as shown in the figure. 
The amplitude of the \Alfven waves is amplified through the upward 
propagation, as a result, a non-linear term due to  
variation of magnetic pressure accompanying with the \Alfven 
waves excites longitudinal slow waves\citep{ks99}. 
They eventually steepen to form MHD slow shocks \citep{suz02} 
which efficiently convert kinetic and magnetic energy to heat. 
The coronal heating and wind acceleration are thus far achieved 
by transfer from the energy and momentum of the outgoing \Alfven waves. 
Linearly polarized \Alfven waves directly 
steepen in themselves to form MHD fast shocks \citep{hol82,suz04}, 
which also contributes to the heating, though it is less dominant.
The incoming \Alfven waves are generated in the corona by the reflection of 
the outgoing ones by the density fluctuations due to the slow waves. 
The reflected waves further play a role in the dissipation of the outgoing 
\Alfven waves by nonlinear wave-wave interaction.

\section{Summary and Discussions}
We have performed one-dimensional MHD simulation for the low-frequency 
\Alfven waves excited at the photosphere. We incorporate radiative cooling 
and thermal conduction, and self-consistently treat the mass, momentum, and 
energy transfer by nonlinearly and dynamically solving the wave propagation 
and dissipation. The advantage of our simulation is the {\em forward} 
approach with the {\em minimal} assumptions. 

Our result has manifestly shown that the formation of the corona and the 
acceleration of the fast solar wind in the coronal holes are the natural 
outcome of the footpoint motions of the field line.     
The photospheric fluctuations generate the low-frequency \Alfven waves which 
propagate upwardly. About $15\%$ of the initial \Alfven waves in energy flux 
can transmit into the corona after surviving the non-WKB reflection in the 
chromosphere and the transition region. The amplitudes of these out-going 
waves are amplified due to the density stratification so that they effectively 
dissipate by the nonlinear effects and heat and accelerate the coronal plasma.

A key mechanism in the coronal heating and the solar wind acceleration is the 
generation of the MHD slow wave. 
Thus, one of the predictions from our simulation is the existence of the 
longitudinal fluctuations in the solar wind plasma.  
This is directly testable by future missions, Solar Orbiter and Solar Probe, 
which will approach to $\sim$45 and 4 $R_{\rm S}$, respectively, 
being in our simulation region.  
They can determine how much fractions of the fluctuations are in the 
transverse and longitudinal modes by in situ measurements which are directly 
compared with our result.

In this paper we have considered the wave propagation and dissipation in 
one-dimensional MHD approximation. However, its validity needs to be examined 
by future studies. It is important to investigate the multidimensional 
effects for the waves \citep{ofm04}, such as refraction \citep{bog03}, 
phase mixing\citep{hp83,nrm98}, turbulent cascade into the transverse 
direction \citep{oug01,dmi02}.  
The waves could suffer collisionless damping (e.g. Suzuki et al.2005), 
and this mechanism might also modify the dissipation rate.  
Kinetic effects of the multicomponent plasma 
might be also important especially in the outer region with low density. 

We need to check how much the wave dissipation and the consequent plasma 
heating are modified by these processes. 
However, we cannot study them in the global simulations as performed here 
because of the huge density difference. 
Hence, we should firstly 
investigate the detailed processes in the 
local simulations, and then, compare with global results;  
the global simulation in the one-dimension and the local simulations in 
the two- or three-dimensions should play complimentary roles.

\section*{Acknowledgment}
We thank Drs. K. Shibata and T. Sano for many fruitful discussions. 
T.K.S. is supported by the JSPS Research Fellowship for Young
Scientists, grant 4607.
This work is supported by the Grant-in-Aid for the 21st Century COE 
"Center for Diversity and Universality in Physics" at Kyoto University 
from the Ministry of Education, Culture, Sports, Science and Technology 
(MEXT) of Japan.

\end{document}